\documentclass[conference]{IEEEtran}
\ifCLASSINFOpdf
\else
\fi
\usepackage{url}

\usepackage{algorithm}
\usepackage{algorithmicx}
\usepackage{algpseudocode}
\usepackage{booktabs} 
\usepackage{amsmath}
\usepackage{breqn}
\usepackage{color}
\usepackage{cleveref} 
\usepackage{graphicx}
\usepackage{caption}
\usepackage[flushleft]{threeparttable}
\usepackage[export]{adjustbox}
\graphicspath{ {./png/}  }

\begin{document}
%
\title{Globalness Detection in Online Social Network}



\author{\IEEEauthorblockN
{Yu-Cheng Lin,
Chun-Ming Lai, 
S. Felix Wu and
George A. Barnett}
University of California, Davis\\
Email: \{ycjlin,cmlai,sfwu,gabarnett\}@ucdavis.edu}


\maketitle

\begin{abstract}
  	Classification problems have made significant progress due to the maturity of artificial intelligence (AI). However, differentiating items from categories without noticeable boundaries is still a huge challenge for machines -- which is also crucial for machines to be intelligent. 
In order to study the fuzzy concept on classification, we define and propose a globalness detection with the four-stage operational flow. We then demonstrate our framework on Facebook public pages inter-like graph with their geo-location. Our prediction algorithm achieves high precision (89\%) and recall (88\%) of local pages. We evaluate the results on both states and countries level, finding that the global node ratios are relatively high in those states (NY, CA) having large and international cities. Several global nodes examples have also been shown and studied in this paper.
It is our hope that our results unveil the perfect value from every classification problem and provide a better understanding of global and local nodes in Online Social Networks (OSNs).

\end{abstract}

%
\IEEEpeerreviewmaketitle

\section{Introduction}
Thanks to the development of graphics processing unit (GPU), artificial intelligence (AI) and machine learning (ML) have achieved great advances in this decade. Many research fields have made great strides, especially in quantitative and cognitive aspects with the efficiency and effectiveness provided by AI and ML.
However, most research problems in social informatics addressing human cognition and behaviors are not applicable with category identification, which has been a critical topic in machine learning. In most cases, it is relatively challenging and arbitrary to classify an item into just a single category since there exist vague items which could be classified as neither category A nor B. For example, discussion messages in social media can be conservative, liberal, or impartial -- which means these messages are not polarized enough that they can be labeled as either liberal or conservative \cite{zafar2016message}.

\begin{figure}
	\centering
	\includegraphics[width=.7\columnwidth]{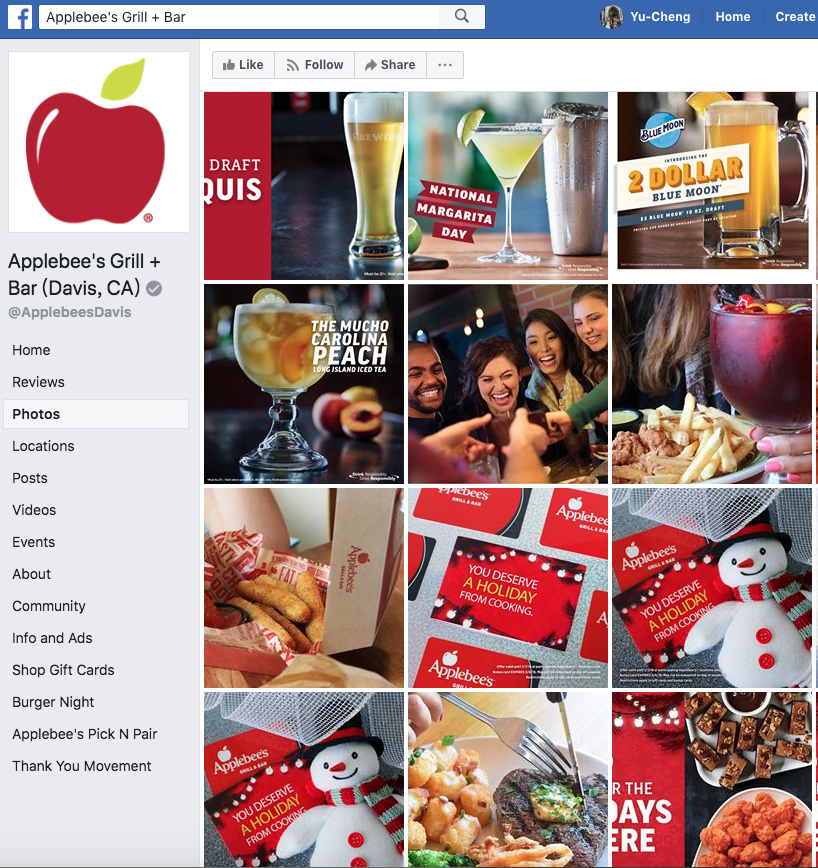}
    \caption{The restaurant chain Applebee's fan page at Davis, CA}
	\label{fig:applebee_davis_fb}
\end{figure}

Conversely, another special case is that an item could be classified into one of the multiple categories -- all seem to be reasonable. For instance, a restaurant page (e.g. Applebee at Davis, displayed in \Cref{fig:applebee_davis_fb}) could be classified as an American-style restaurant, bar, grill restaurant, and even a steak restaurant at the same time. In general, most labelled tasks require manual feature selection and interpretation, which facilitates the recent popularity of deep and reinforcement learning. In the learning process, each item is given a combination of probability distribution of all possible categories \cite{lecun2015deep}. Nonetheless, we still need to determine the classification results by the category with the highest probability ultimately. 

We seek to complement this fundamental problem of classification by proposing the framework of globalness detection -- \textit{finding the items which does not belong to a specific category.}
This framework is especially useful in social science field since human perception is not just a simple binary classification problem
. Sometimes we are confident to perform near-perfect algorithms and get results with large-scale data training and validation. For example, in pattern recognition, dogs should not be identified as birds because dogs have tails and four legs while birds have wings and two legs -- the distinction is clear among different categories.
Instead, in most cases when considering social informatics research problems, there are usually no clear definition and support vectors to build a convincing classifier.  

In this paper, we try to approach this classification challenge by uncovering the global nodes in OSNs. We define global nodes as those nodes, instead of being local, having an unbiased relationship with the nodes in multiple categories.
Our objective is to build an accurate classifier which can identify both local and global nodes with high precision. 
In our methodology, we employ anchor nodes as the center of localization to the corresponding categories, which is similar to the idea applied in sensor networks \cite{akl2011anchor}. The detailed operational flow is described in \Cref{sec:operationalizing}.

Globalness detection in OSNs (e.g. Facebook) has two directions of potential applications. First, understanding the location of your target audience is a primary tool in the marketing strategy toolbox. In this way, locality differentiation between various Facebook pages is crucial to ensuring that marketers can understand their audiences as discrete geo-located elements. For instance, BMW has distinct advertisements based on elements such as geography, culture, and language \cite{polavskova2013representation}. 
Another example involves media broadcasters; CBS News has divided its viewing audiences into separate groups based on geographic data. They exploit this classification as a virtual demarcation line in order to broadcast different messages to different communities \cite{althaus2009media}.
In short, locality differentiation facilitates the subtle division of information for multiple audiences.

Second, there are always positive and negative supporters for a specific standing in OSNs. It's relatively accessible to find these polarized group members from the hybrid approach combining content and networks features \cite{lu2015biaswatch, wong2016quantifying}. However, from the information entropy's point of view, the neutral group members or weakly supporting group members provide much more valuable arguments. Our novel methodology offers a new avenue into this open research problem.

The rest of this paper is organized as the following. 
In \Cref{sec:operationalizing}, the concept of globalness is characterized and defined. Then we presents the proposed operational flow to discover global nodes.
\Cref{sec:data} provides a detailed description for our dataset. Two experimental cases are studied in \Cref{sec:case_study} to illustrate the power of our platform.
In \Cref{sec:related}, we summarize past research on node analysis in OSN and data collection of location information.
Finally, \Cref{sec:conclu} offers a summary of our work.

\section{Operationalizing Globalness}
\label{sec:operationalizing}
For the computational social science, it's important to operationalize an intuitively fuzzy notion -- making it quantitatively distinguishable in terms of empirical observations. Since globalness is not a measurable variable, determining an appropriate approach is an important part of the study. Specifically, we try to answer these two research questions:

\textbf{RQ1}: \textit{What does it mean for an item (node) to be global in online social network graph?}

\textbf{RQ2}: \textit{How can we detect these global nodes in online social network graph?}

In the following, we start from an informal characterization and then propose an innovative method to operationalize this phenomenon. With the methodology of the operational flow, we are able to detect these global nodes in large-scale.

\subsection{Characterizing Concept of Globalness}
\begin{itemize}
    \item Globalness is a \textit{property of the relationship of this node with not only the category it belongs to, but also with other categories.} This follows the intuition that comparing to the local nodes, the global nodes are not inclined towards a few categories.
    \item Globalness \textit{is not an absolute concept and is defined with respect to a given context:} the nodes of a graph and a given set of (two or more) parties. For instance, a Facebook page could be global from Midwest's point of view, but could be local from the perspective of the United States (US).
\end{itemize}

\subsection{Operational Definition of Globalness}
From our observation, the global nodes in OSNs are those nodes which have connections with multiple categories (classes or regions) and do not lean toward a specific category.
For instance, given all the Facebook pages located within the US, some of them have characteristics which are similar to the pages in the other countries. 
We propose an operational definition of this concept by the following formula:
\begin{equation}
\begin{split}
Global\_node := \{ p \space | p \space \in C_{i}{} \space \land 
 ( \sum_{j \in C}^{} \delta_{pj} ) >= N_{global\_threshold} \} \\
\end{split}
\end{equation}

\begin{equation}
\begin{split}
\delta_{pj} = 
    \begin{cases}
      1, & \Delta_{pj} \leq \min_{k \in C} (\Delta_{pk}) + \epsilon \\
      0, & \text{otherwise}
    \end{cases}
\end{split}
\end{equation}

\begin{equation}
\begin{split}
\Delta_{pk} = \sum_{c \in C \space \land \space c \neq k}^{} \omega_{c}|| D_{pc} ||
\end{split}
\end{equation}

where $p$ is the global node which originates from the class $C_{i}{}$; $\Delta_{pk}$ is the weighted error when node $p$ is classified as the class $C_{k}{}$ and $D_{pc}{}$ is the error distance to a specific class.
In short, abstractly global nodes are those nodes which could be classified as anyone of the multiple classes within a small error distance.
\Cref{fig:global_page_flow} demonstrates an example flow to detect global pages in the US.

\subsection{Operational Flow of Globalness Detection}
\label{operational_flow_globalness}
In order to find the global nodes in OSNs, we devise a four-stage operational flow of globalness detection.
\subsubsection{Making hypothesis} Define globalness relationship between two affiliations. Specifically, there are mainly two groups: A and non-A (others).
Note that while this definition can be intuitively extended to relationship among more than two affiliations, for the sake of simplicity, we only consider relationship between two affiliations in this work. 

\subsubsection{Deciding anchor nodes} For each class, pick one anchor node which can exemplify its own class -- namely, choose the representative "local" nodes at this stage.

\subsubsection{Selecting biased data} In the training process, we reinforce the distinction between local nodes and global nodes by polarizing the selection of training samples. For the local nodes, we choose the nodes with local tendency in class A; for the global nodes, we pick the nodes with global propensity in class non-A.

\subsubsection{Detecting global nodes by classification} At this stage, we train the model by the biased data. We identify these global nodes by the trained classifier. The global nodes are the nodes which are identified as belonging to the class "others (OT)" by the classifier -- which means that these nodes do not belong to their original class, and thus are so-called \textit{global nodes} by our operational definition. 

\begin{figure} 
	\centering
	\includegraphics[width=0.8\columnwidth]{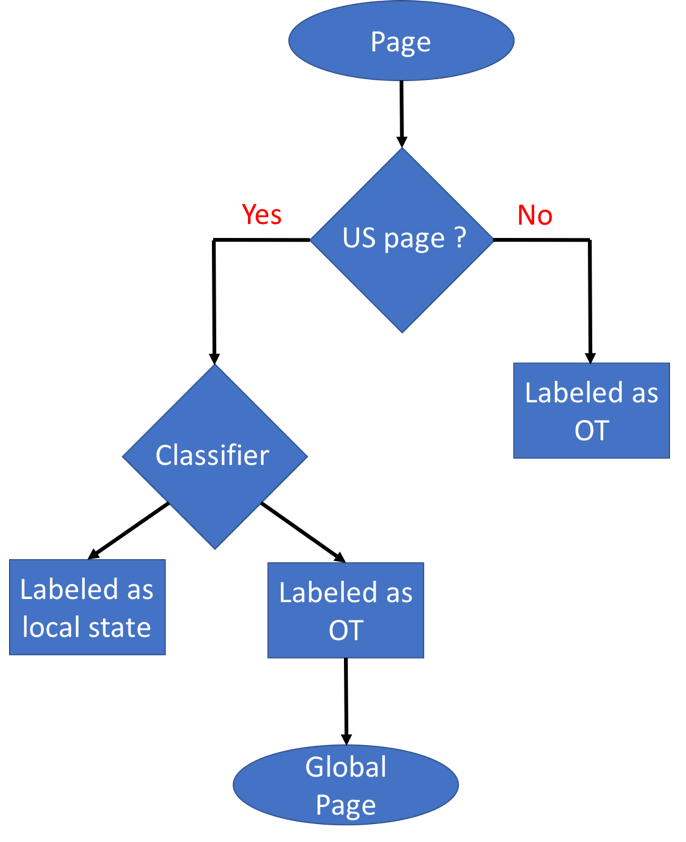}
	\caption{Global page detection flow for the US pages}
	\label{fig:global_page_flow}
\end{figure}

\section{Dataset}
\label{sec:data}
In this paper, our experiments from the operational flow of global detection are materialized on the Facebook public pages.

\subsection{Facebook Public Page Graph}
Formally, we model the Facebook public pages
as a page-like graph $G=(V,E)$ wherein V is a set of pages and E is a set of directed edges presenting "Like" relationship which would be seen by the community participants.

The page host has the privilege to disclose the
geographic location of the page. We crawled
38,831,367 pages in total. 
Among these pages, 15,115,963 (38.93\%) pages have specific location information. 
There are 2,430,873 pages proclaimed as in the United States (US) and 12,685,090 non-US pages.
We took the pages with declared location information of country and city as ground truth data. 
Then we map the city to the corresponding state where it resides. Few pages are excluded because their city names exist in multiple states, which can result in ambiguous city-to-state mapping. There are 29,849 cities in total in the US.

\subsection{Anchor Node Selection}
\label{anchor_node_selection}
Seed selection is an important technique in graph analysis and also exploited in combating fake contents (e.g. web spam and review manipulation) \cite{kaghazgaran2018combating, gyongyi2004combating}.

In \Cref{operational_flow_globalness}, the second step of the operational flow is to decide anchor nodes representative of their own categories. These anchor nodes have to be as local as possible such that the distances to these nodes can afford the authentic tendency towards locality.

In our dataset, we select all subsidiary pages of "OnlyInYourState.com" as the anchor nodes. For example, "Only\_In\_Delaware" is part of "OnlyInYourState.com" and mostly connects the local communities in Delaware.
In addition, since California is much larger than the other states in the perspectives of population and economy, "OnlyInYourState.com" splits California into Northern and Southern regions.
As a result, both
"Only\_In\_Northern\_California" and "Only\_In\_Southern\_California" are used as anchor
nodes. Furthermore, since "Only\_In\_Idaho" had been registered, OnlyInYourState.com named its Idaho counterpart as "Idaho\_Only" instead. In sum, we have 51 anchor nodes in total for this dataset.

\section{Case Study: Globalness in Facebook Pages}
\label{sec:case_study}
To illustrate the methodology we proposed to detect global nodes in OSN, in this section we investigate two cases utilizing Facebook public pages as the graph nodes.

\subsection{Implementation on Facebook Public Pages}
With the anchor node as the center of each state, we adopt the breadth-first search (BFS) to calculate the distance from each anchor node.
First, we define the state distance vector (SDV) to represent the hop distances, for each specific page, to the anchor pages.
The following equation shows SDV for each page:
\begin{equation}
\begin{split}
 SDV(Page) = [\qquad \qquad \qquad \qquad \\
 [IHOP(Page, S_{i}), \qquad \qquad\\
 OHOP(Page, S_{i})]: \qquad \quad\\
 i \in {1, ..., N_{anchored\_pages}}]
\end{split}
\end{equation}
where $IHOP(Page, S_{i})$ denotes the hop distance between the page and the seed $S_{i}$, using the inward edges as connection for BFS; $OHOP(Page, S_{i})$ denotes the hop distance between the page and the seed $S_{i}$, using the outward edges as connection for BFS. 
In addition, we define the minimum hop distance $MHOP$ as the minimum of all hop distances of a page:
\begin{equation}
\begin{split}
 MHOP = Min(Min(IHOP), Min(OHOP))
\end{split}
\end{equation}
where $IHOP$ is the set of $IHOP(Page, S_{i})$ and $OHOP$ is the set of $OHOP(Page, S_{i})$.

From our observation, in some cases, an anchor page can
connect with those pages far beyond where it resides within a small hop distance.
This probability is not high; however, this kind of connection noise would highly reduce the prediction accuracy.
To address this problem, we take the neighborhood location probability into consideration as the auxiliary features. The state neighborhood probability (SNP) presents the location information of the adjacent pages of a specific page. 
The definition of SNP is shown below:
\begin{equation}
\begin{split}
 SNP(Page) = [\qquad \qquad \qquad \qquad \\
 [INP(Page, R_{i}), \qquad \qquad\\
 ONP(Page, R_{i})]: \qquad \quad\\
 i \in {1, ..., N_{region}}]\qquad \quad
\end{split}
\end{equation}
where $INP(Page, R_{i})$ denotes the inward neighborhood location probability between this page and the adjacent pages belonging to the region $R_{i}$; $ONP(Page, R_{i})$ denotes the outward neighborhood location probability between the page and the adjacent pages belonging to the region $R_{i}$. 
The equations are as the following:
\begin{equation}
\begin{split}
 INP(Page, R_{i}) = \dfrac{IE(Page, R_{i})}{\sum_{i}^{}IE(Page, R_{i})}\\
 ONP(Page, R_{i}) = \dfrac{OE(Page, R_{i})}{\sum_{i}^{}OE(Page, R_{i})}\\
\end{split}
\end{equation}
where $IE(Page, R_{i})$ is the number of the inward edges between this page and the adjacent pages belonging to the region $R_{i}$; $OE(Page, R_{i})$ is the number of the outward edges between this page and the adjacent pages belonging to the region $R_{i}$.

\begin{figure}
	\centering
	\includegraphics[width=.9\columnwidth]{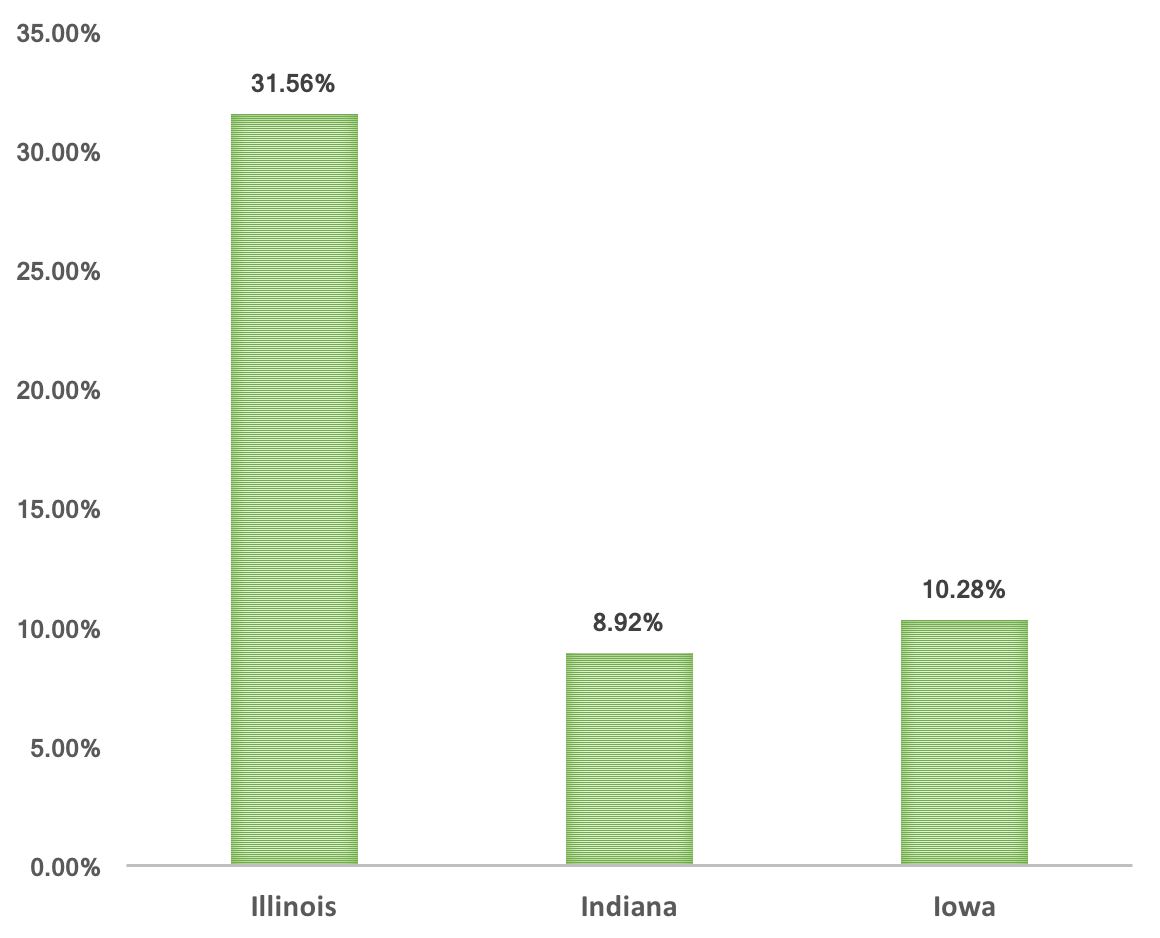}
    \caption{Comparison of state-wise global page percentage}
	\label{fig:global_page_state_wise}
\end{figure}

\subsection{State-wise Globalness Detection}
\label{state_wise_experiment}
To investigate the global nodes in the state-wise perspective, we select Facebook pages in the threes states of the Midwest region: Illinois (IL), Indiana (IN), Iowa (IA).

\subsubsection{Making hypothesis} For IL globalness detection, we define global pages (OT) in IL as those pages labeled as IL, but behave as those pages which reside in non-IL states (e.g. IN and IA in this experiment setting) and have state-wise global tendency -- which means a balanced relationship with all three states (IL, IN, and IA).

\subsubsection{Deciding anchor nodes} As described in \Cref{anchor_node_selection}, we pick the three anchor nodes in IL, IN, and IA.

\subsubsection{Selecting biased data} We choose the pages labeled as IL and $\#MHOP <= N_{local\_threshold} = 1$ as the local node training samples; we pick the pages labeled as others (i.e. IN and IA) and $\#MHOP >= N_{global\_threshold} = 3$ as the global node training samples.

\subsubsection{Detecting global nodes by classification} The trained classifier is deployed to recognize the global pages in IL. There are two categories in this classifier: IL and 
"others (OT)" representing the global pages.

The same globalness detection flow is applied accordingly to detect global nodes in both IN and IA.
\Cref{fig:global_page_state_wise} compares the global page percentage in the three states (IL, IN, and IA).
31.56\% of the pages in IL are global pages, which are at least 20\% more than the counterparts in IN and IA.
This result suggests that IL has much more pages which actively interact with other pages among the three core states in the Midwest \cite{midwest_states_poll} -- which supports the fact that IL, home of Chicago, is the center of the Midwest region.

\begin{figure}
	\centering
	\includegraphics[width=.9\columnwidth]{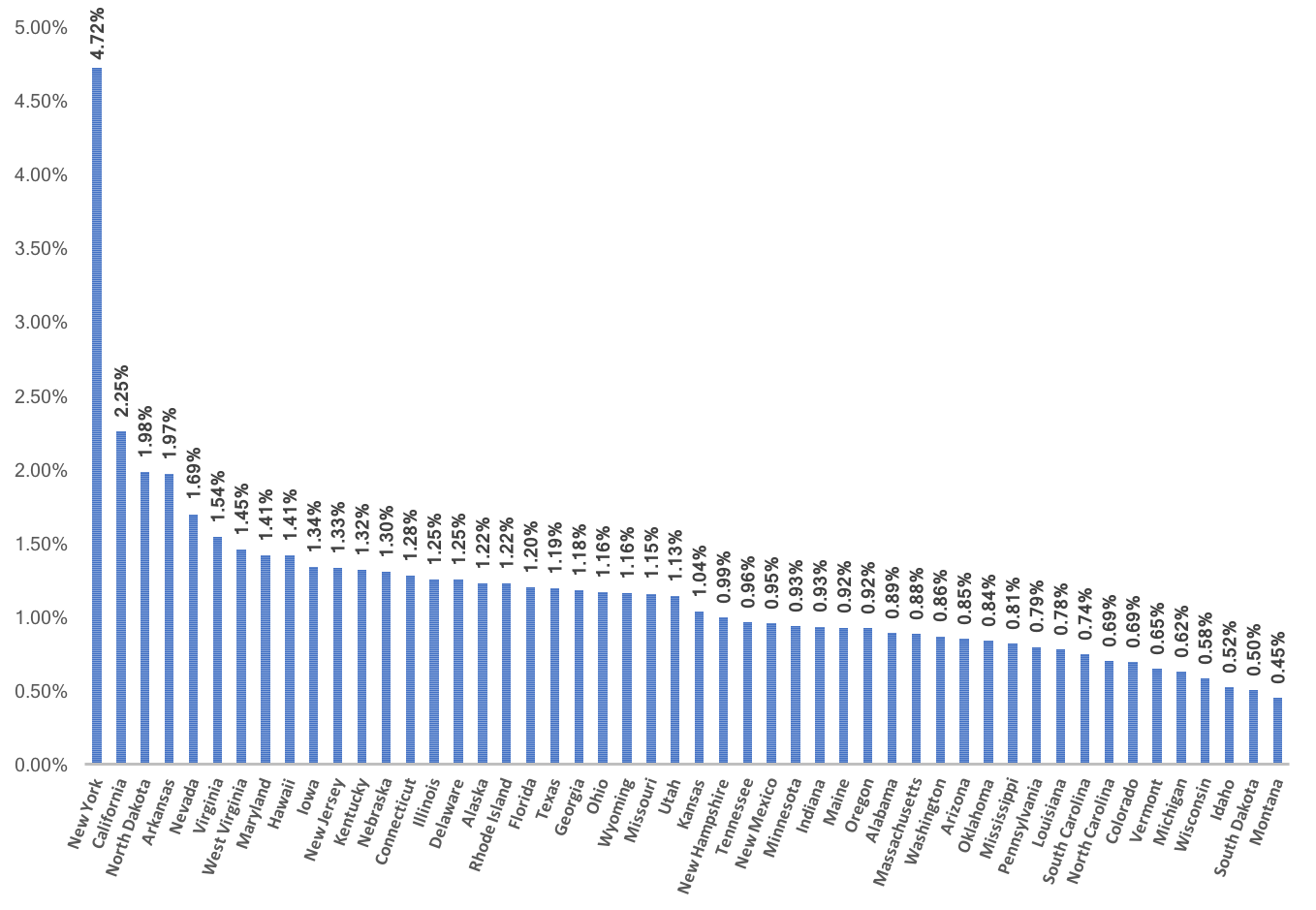}
	\caption{Global page distribution by state}
	\label{fig:global_page_dist}
\end{figure}

\begin{figure}
	\centering
	\includegraphics[width=.9\columnwidth]{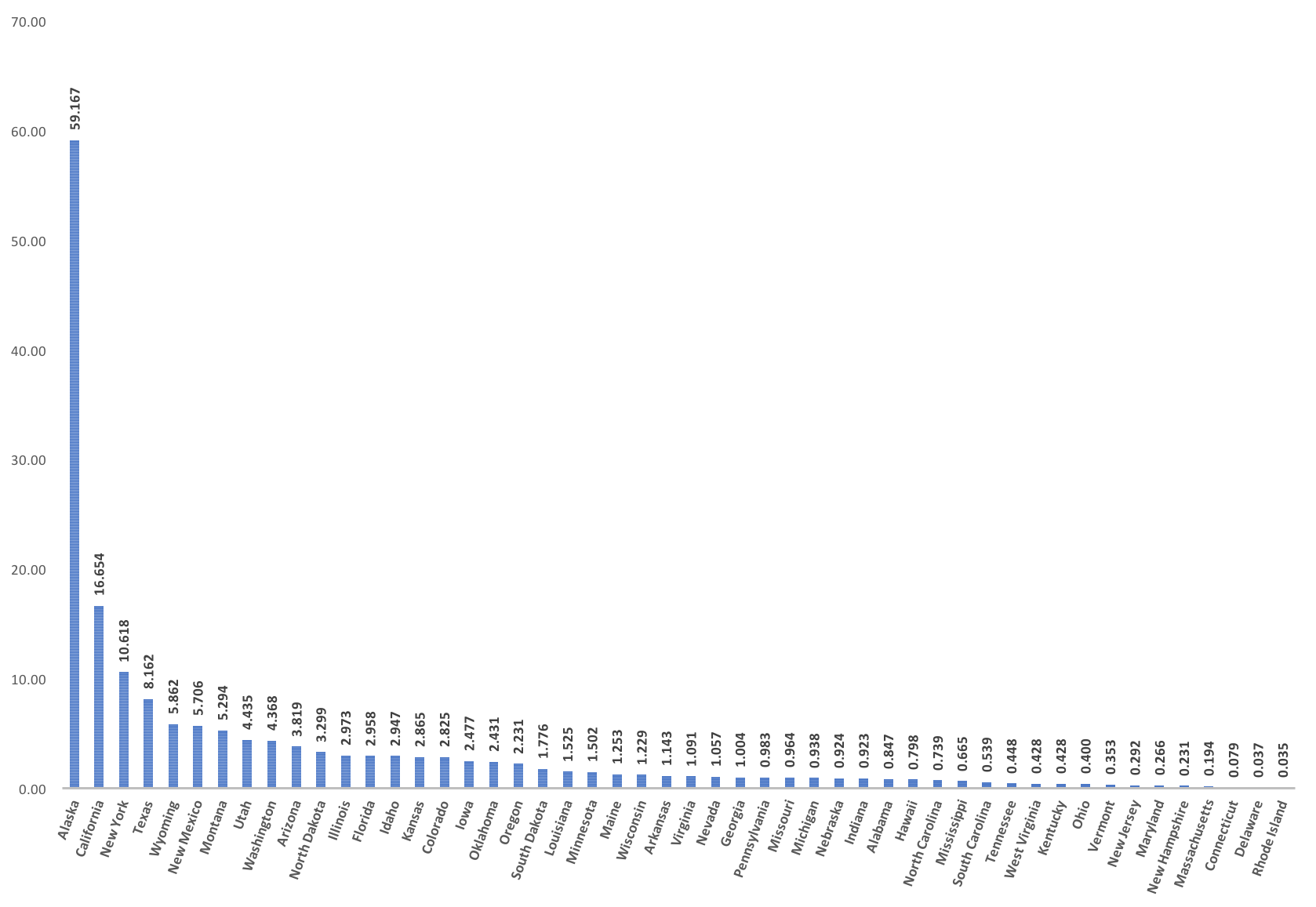}
	\caption{Global page ratio w.r.t population density by state}
	\label{fig:gb_ratio}
\end{figure}

\subsection{Country-wise Globalness Detection}
\label{country_wise_experiment}
In this section, we explore the possibility of applying our methodology onto globalness detection in the country-wise aspects. 
As an example, the US global pages are distinguished in the following steps, as shown in \Cref{fig:global_page_flow}.

\subsubsection{Making hypothesis} We define global pages (OT) in the US as those pages labeled as states in the US, but behave as those pages which are outside the US and have global propensity.

\subsubsection{Deciding anchor nodes} As described in \Cref{anchor_node_selection}, we choose all 51 anchor nodes for the 50 states in the US.

\subsubsection{Selecting biased data} We choose the pages labeled as US and $\#MHOP <= N_{local\_threshold} = 2$ as the local node training samples; we pick the pages labeled as others (non-US) and $\#MHOP >= N_{global\_threshold} = 5$ as the global node training samples.
Both local threshold $N_{local\_threshold}$ and global threshold $N_{global\_threshold}$ could be adjusted to make the biased selection more polarized.

\subsubsection{Detecting global nodes by classification} The trained classifier is deployed to identify the global pages in US. There are 51 categories in this classifier: fifty US states and "others (OT)" representing the global pages.

\Cref{fig:global_page_dist} presents the global page distribution by state.
The proportion of global pages in all states is 1.16\% on average.
Not surprisingly, NY has the highest global page ratio (4.72\%).
This result is consistent with NY's image as a global 'melting pot.'
CA is the other state whose global page ratio is higher than 2\%, which shows its global diversity.
From demographics' point of view, we rank states by the ratio of $(\#Global \enspace pages) \, / \, (Population \enspace density)$ \cite{demographic_stats}.
As \Cref{fig:gb_ratio} shows, AS ranks the 1st (59.2), which manifests its endeavor in travel promotion, and the followings are the most populous states: CA, NY, and TX.

\subsection{Evaluation}
In both cases in the above experiments, we've applied three different ML models: Naive Bayes \cite{zhang2004optimality},
Adaboost \cite{hastie2009multi}, and
Random Forests \cite{breiman2001random}. 
The Random Forests model gets the best performance. Hence we only presents the results based on this model in this paper.

In \Cref{state_wise_experiment}, it's a simplified classification problem, and our classifiers get 100\% precision and recall rates when recognizing local pages in IL, IN, and IA. 
There are 50 state categories to identify in \Cref{country_wise_experiment}, which is a much more realistic and complex experiment. Our classifier gets \textbf{89\%} precision and \textbf{88\%} recall rate. The detailed prediction results by state are provided in \cite{lin2018geo}.

Next, we evaluate the accuracy of globalness detection for our approach. To that end, we performed a simple experiment: We randomly sampled 
15 pages from each state. Then one of the authors manually verified whether these 15 pages can be really classified as global page by human judgment. We found that 100\% of these pages can be classified as global pages, which have connections with pages in multiple other states and behave like pages outside the US.

It is important to highlight that our methodology was not designed to capture all of the global pages. In addition, the parameter settings of the thresholds give flexibility to trade-off between precision and recall for globalness detection. Our approach aimed at building a high precision prediction platform for both local and global nodes.

\begin{figure}
	\centering
	\includegraphics[width=.8\columnwidth]{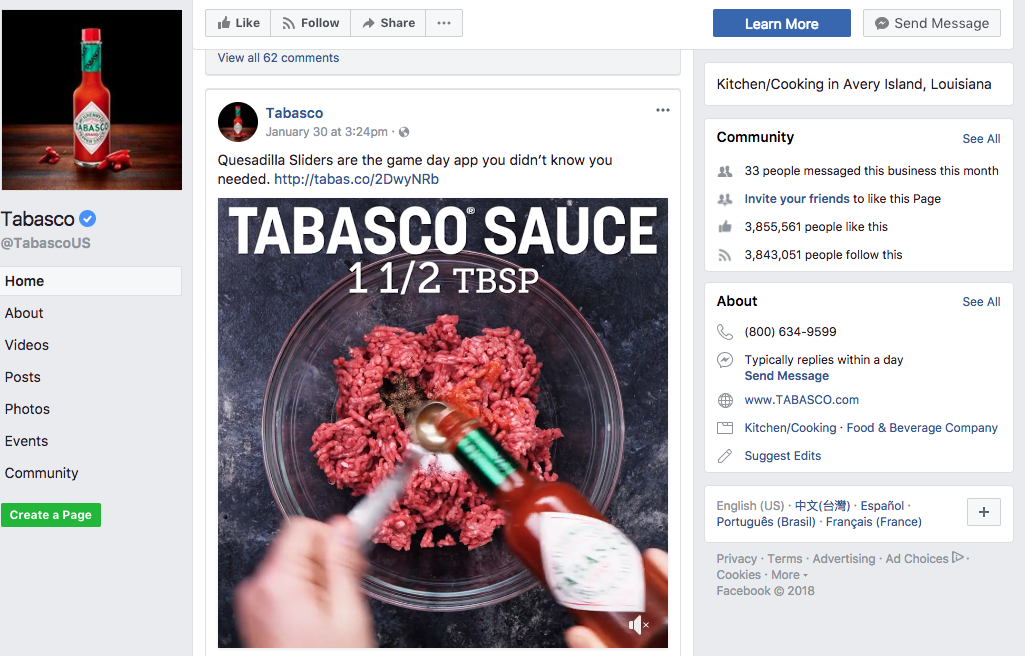}
    \caption{Demonstration of a global page located in LA}
	\label{fig:global_page_demo_inter-state}
\end{figure}

\begin{figure}
	\centering
	\includegraphics[width=.8\columnwidth]{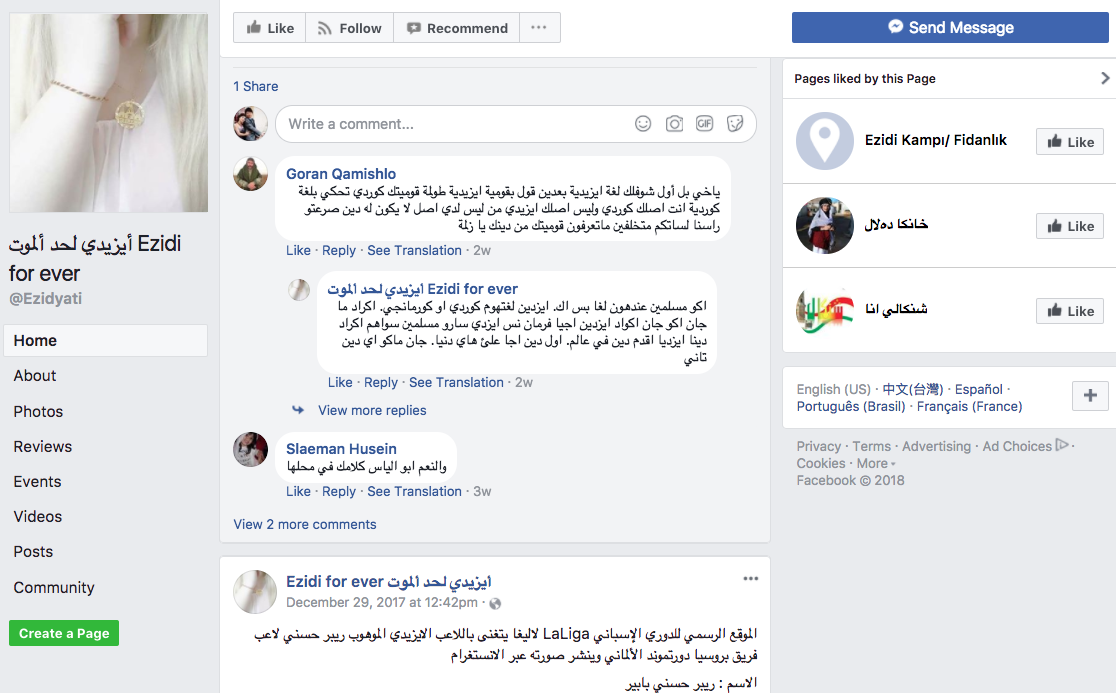}
    \caption{Demonstration of a global page located in NY}
	\label{fig:global_page_demo_international}
\end{figure}

In sum, there are primarily two kinds of global pages: the commercial pages which intend to connect multiple regions for promotion purposes, and the immigrant communities which are closely linked with exotic contents.
For instance, the page shown in \Cref{fig:global_page_demo_inter-state} is Tabasco's fan page; \Cref{fig:global_page_demo_international} 
(page ID = 1627668764132931) is identified as a global page -- located in NY, but highly related to the Middle East culture.

\subsection{Stability}
In \Cref{country_wise_experiment}, our experimental results are based on the anchor nodes of the subsidiary pages from "OnlyInYourState.com." However, from our assumption, the anchor nodes are not unique as long as they are local enough to provide accurate distance measurement in BFS process. Then what's the stability of our globalness detection results?

From our observation, state university pages are also a possible source set of anchor nodes since local students tend to attend the state universities in their residence states. Therefore, to make a fair comparison, we pick one state university page for Northern California, one state university page for Southern California, and one state university page for the other states.
Our classifier gets 88\% precision and 88\% recall rate. These comparable results suggest that the anchor set of state university pages is also a valid anchor set.

From the perspective of platform stability, we compare the global pages distinguished from the two sets of anchor nodes. In sum, there are around 60\% global pages overlapped between these two sets. This high repetition rate proves the stability of our globalness detection platform with distinct anchor sets.

\section{Related Work}
\label{sec:related}
\subsection{Node Analysis in Online Social Network}
In OSN, users, pages, and posts/comments are nodes comprising a multi-layer composite social network graph. Most research papers analyze the nodes in a single layer, with another layer as a supplement.
Firstly, Ugander \emph{et al.} analyzed the Facebook user graph from network structure properties \cite{ugander2011anatomy}. 
Second, Hong \emph{et al.} 
focused on the analysis of the Facebook public page graph with the data-driven methodology \cite{hong2018profiling}. 
Last but not least, the political polarity of posts/comments in OSN (e.g. Twitter) is also a popular topic many researchers have been working on \cite{zafar2016message, lu2015biaswatch, wong2016quantifying, speriosu2011twitter, maynard2011automatic}.

\subsection{Location Information}
The ground truth location information plays a key role in model building and evaluation of geo-location prediction. There are two methodologies to acquire this information: direct collection and indirect extraction.

First, direct collection relies on the OSN platform to 
fetch data. With a non-required text field, guided in the case of Facebook and freeform in the case of Twitter, both services provide users with the option to proclaim their locations. This field can facilitate the generation of location labels from country, state, to city level.
According to Hecht's report, 34\% of users provide invalid geographic information in this location field on Twitter \cite{hecht2011tweets}. Another more accurate way is to adopt the coordinate location instead of the named location \cite{cheng2010you}. 

Second, indirect extraction uses crowdsourcing to tag the data. Huang \emph{et al.} adopted a composite approach with 
the language detection tools, an R library, as well as the Yahoo! and Twitter API to make the crowdsourcing job easier for people to label data \cite{huang2014inferring}.

Since the majority of public pages are commercial in nature, and as such are managed by page hosts who have a desire to make these pages thriving, they tend to use more accurate location data in order to ensure more viable commercial traffic \cite{hong2018profiling}. As users have no such intrinsic motivation to provide accurate location data, this means the quality of location attribute data is much higher for pages than for users. For this reason we used direct collection to obtain location data on Facebook.

\section{Conclusion}
\label{sec:conclu}
In this paper, we first characterized and defined the concept of globalness to answer \textbf{RQ1}: \textit{What does it mean for an item (node) to be global in online social network graph?} Secondly, an operational flow of globalness detection was introduced to answer \textbf{RQ2}: \textit{How can we detect these global nodes in online social network graph?} Based on this flow, we examined two cases to find global nodes in Facebook pages. 
Our results are encouraging and reasonable, indicating that this framework can indeed uncover global nodes in the OSN graph.
For future work, we are also eager to further explore 
other possible applications including social media's political standing (e.g. liberal, conservative, or neutral), from the perspective of relationship.
In addition, since one primary source of global pages is related to commercial promotion, the judgement of globalness could also be taken as a feature of malicious link detection in Facebook pages \cite{lai2017attacking}.

\bibliographystyle{IEEEtran}
\bibliography{references}

\begin{thebibliography}{10}
\providecommand{\url}[1]{#1}
\csname url@samestyle\endcsname
\providecommand{\newblock}{\relax}
\providecommand{\bibinfo}[2]{#2}
\providecommand{\BIBentrySTDinterwordspacing}{\spaceskip=0pt\relax}
\providecommand{\BIBentryALTinterwordstretchfactor}{4}
\providecommand{\BIBentryALTinterwordspacing}{\spaceskip=\fontdimen2\font plus
\BIBentryALTinterwordstretchfactor\fontdimen3\font minus
  \fontdimen4\font\relax}
\providecommand{\BIBforeignlanguage}[2]{{%
\expandafter\ifx\csname l@#1\endcsname\relax
\typeout{** WARNING: IEEEtran.bst: No hyphenation pattern has been}%
\typeout{** loaded for the language `#1'. Using the pattern for}%
\typeout{** the default language instead.}%
\else
\language=\csname l@#1\endcsname
\fi
#2}}
\providecommand{\BIBdecl}{\relax}
\BIBdecl

\bibitem{zafar2016message}
M.~B. Zafar, K.~P. Gummadi, and C.~Danescu-Niculescu-Mizil, ``Message
  impartiality in social media discussions.'' in \emph{ICWSM}, 2016, pp.
  466--475.

\bibitem{lecun2015deep}
Y.~LeCun, Y.~Bengio, and G.~Hinton, ``Deep learning,'' \emph{Nature}, vol. 521,
  no. 7553, pp. 436--444, 2015.

\bibitem{akl2011anchor}
R.~Akl, K.~Pasupathy, and M.~Haidar, ``Anchor nodes placement for effective
  passive localization,'' in \emph{Mobile and Wireless Networking (iCOST), 2011
  International Conference on Selected Topics in}.\hskip 1em plus 0.5em minus
  0.4em\relax IEEE, 2011, pp. 127--132.

\bibitem{polavskova2013representation}
T.~Pol{\'a}{\v{s}}kov{\'a}, ``The representation of luxury products in printed
  advertisements,'' 2013.

\bibitem{althaus2009media}
S.~L. Althaus, A.~M. Cizmar, and J.~G. Gimpel, ``Media supply, audience demand,
  and the geography of news consumption in the united states,'' \emph{Political
  Communication}, vol.~26, no.~3, pp. 249--277, 2009.

\bibitem{lu2015biaswatch}
H.~Lu, J.~Caverlee, and W.~Niu, ``Biaswatch: A lightweight system for
  discovering and tracking topic-sensitive opinion bias in social media,'' in
  \emph{Proceedings of the 24th ACM International on Conference on Information
  and Knowledge Management}.\hskip 1em plus 0.5em minus 0.4em\relax ACM, 2015,
  pp. 213--222.

\bibitem{wong2016quantifying}
F.~M.~F. Wong, C.~W. Tan, S.~Sen, and M.~Chiang, ``Quantifying political
  leaning from tweets, retweets, and retweeters,'' \emph{IEEE transactions on
  knowledge and data engineering}, vol.~28, no.~8, pp. 2158--2172, 2016.

\bibitem{kaghazgaran2018combating}
P.~Kaghazgaran, J.~Caverlee, and A.~Squicciarini, ``Combating crowdsourced
  review manipulators: A neighborhood-based approach,'' in \emph{Proceedings of
  the Eleventh ACM International Conference on Web Search and Data
  Mining}.\hskip 1em plus 0.5em minus 0.4em\relax ACM, 2018, pp. 306--314.

\bibitem{gyongyi2004combating}
Z.~Gy{\"o}ngyi, H.~Garcia-Molina, and J.~Pedersen, ``Combating web spam with
  trustrank,'' in \emph{Proceedings of the Thirtieth international conference
  on Very large data bases-Volume 30}.\hskip 1em plus 0.5em minus 0.4em\relax
  VLDB Endowment, 2004, pp. 576--587.

\bibitem{midwest_states_poll}
FiveThirtyEight, ``Which states are in the midwest?''
  https://fivethirtyeight.com/features/what-states-are-in-the-midwest, 2014.

\bibitem{demographic_stats}
ipl2, ``States ranked by size \& population,''
  http://www.ipl.org/div/stateknow/popchart.html, 2012.

\bibitem{zhang2004optimality}
H.~Zhang, ``The optimality of naive bayes,'' \emph{AA}, vol.~1, no.~2, p.~3,
  2004.

\bibitem{hastie2009multi}
T.~Hastie, S.~Rosset, J.~Zhu, and H.~Zou, ``Multi-class adaboost,''
  \emph{Statistics and its Interface}, vol.~2, no.~3, pp. 349--360, 2009.

\bibitem{breiman2001random}
L.~Breiman, ``Random forests,'' \emph{Machine learning}, vol.~45, no.~1, pp.
  5--32, 2001.

\bibitem{lin2018geo}
Y.-C. Lin, C.-M. Lai, J.~W. Chapman, S.~F. Wu, and G.~A. Barnett,
  ``Geo-location identification of facebook pages,'' in \emph{Proceedings of
  the 2018 International Conference on Advances in Social Networks Analysis and
  Mining (ASONAM 2018)}.\hskip 1em plus 0.5em minus 0.4em\relax IEEE Computer
  Society, 2018.

\bibitem{ugander2011anatomy}
J.~Ugander, B.~Karrer, L.~Backstrom, and C.~Marlow, ``The anatomy of the
  facebook social graph,'' \emph{arXiv preprint arXiv:1111.4503}, 2011.

\bibitem{hong2018profiling}
Y.~Hong, Y.-C. Lin, C.-M. Lai, S.~F. Wu, and G.~A. Barnett, ``Profiling
  facebook public page graph,'' in \emph{Computing, Networking and
  Communications (ICNC), 2018 International Conference on}.\hskip 1em plus
  0.5em minus 0.4em\relax IEEE, 2018.

\bibitem{speriosu2011twitter}
M.~Speriosu, N.~Sudan, S.~Upadhyay, and J.~Baldridge, ``Twitter polarity
  classification with label propagation over lexical links and the follower
  graph,'' in \emph{Proceedings of the First workshop on Unsupervised Learning
  in NLP}.\hskip 1em plus 0.5em minus 0.4em\relax Association for Computational
  Linguistics, 2011, pp. 53--63.

\bibitem{maynard2011automatic}
D.~Maynard and A.~Funk, ``Automatic detection of political opinions in
  tweets,'' in \emph{Extended Semantic Web Conference}.\hskip 1em plus 0.5em
  minus 0.4em\relax Springer, 2011, pp. 88--99.

\bibitem{hecht2011tweets}
B.~Hecht, L.~Hong, B.~Suh, and E.~H. Chi, ``Tweets from justin bieber's heart:
  the dynamics of the location field in user profiles,'' in \emph{Proceedings
  of the SIGCHI conference on human factors in computing systems}.\hskip 1em
  plus 0.5em minus 0.4em\relax ACM, 2011, pp. 237--246.

\bibitem{cheng2010you}
Z.~Cheng, J.~Caverlee, and K.~Lee, ``You are where you tweet: a content-based
  approach to geo-locating twitter users,'' in \emph{Proceedings of the 19th
  ACM international conference on Information and knowledge management}.\hskip
  1em plus 0.5em minus 0.4em\relax ACM, 2010, pp. 759--768.

\bibitem{huang2014inferring}
W.~Huang, I.~Weber, and S.~Vieweg, ``Inferring nationalities of twitter users
  and studying inter-national linking,'' in \emph{Proceedings of the 25th ACM
  conference on Hypertext and social media}.\hskip 1em plus 0.5em minus
  0.4em\relax ACM, 2014, pp. 237--242.

\bibitem{lai2017attacking}
C.-M. Lai, X.~Wang, Y.~Hong, Y.-C. Lin, S.~F. Wu, P.~McDaniel, and H.~Cam,
  ``Attacking strategies and temporal analysis involving facebook discussion
  groups,'' in \emph{Network and Service Management (CNSM), 2017 13th
  International Conference on}.\hskip 1em plus 0.5em minus 0.4em\relax IEEE,
  2017, pp. 1--9.

\end{thebibliography}


\end{document}